# Quantum optical synthesis of high-dimensional ultrafast frequency-bin qudits

**Prasad Koviri[1,†], Tomoya Okita[1,†], Rina Yabumoto[1], Yuta Fujihashi[1], Masahiro Yabuno[2], Hirotaka Terai[2], Shigehito Miki[2], Kali P. Nayak[1] & Ryosuke Shimizu[1,3*]**

[1]*Graduate School of Informatics and Engineering, The University of Electro-Communications, 1-5-1 Chofugaoka, Chofu, Tokyo 182-8585, Japan*

[2]*Advanced ICT Research Institute, National Institute of Information and Communications Technology, 588-2 Iwaoka, Nishi-ku, Kobe, Hyogo 651-2492, Japan*

[3]*Institute for Advanced Science, The University of Electro-Communications, 1-5-1 Chofugaoka, Chofu, Tokyo 182-8585, Japan*

[†]*These authors contributed equally to this work.*

**r-simizu@uec.ac.jp*

**Abstract:** Frequency modes of light are one of the most promising platforms that provide access to high-dimensional quantum states amongst different photonic degrees of freedom capable of high-dimensionality, enabling robust, error-tolerant, and scalable quantum optical information systems. We demonstrate engineering of precisely controlled two-photon high-dimensional states entangled in frequency through time-domain Fourier optical synthesis. We generate and convert a continuous broadband frequency-entangled state into a large range of discrete frequency bins suitable for ITU standards, with spacings ranging from 12.5 GHz to 750 GHz, and observe spectral anticorrelations over 38 frequency bins, including intra-bin pure states at a 100 GHz bin spacing. We characterize the full quantum state dimensionality via Schmidt decomposition and observe lower bounds on the frequency-binned Hilbert-space dimensionalities of at least 289, formed by two entangled qudits with dimension 17. Furthermore, we demonstrate quantum nonlocality via frequency correlations in a transmission experiment over a campus-scale two-node fiber network. This work represents a crucial step towards building a versatile and relatively simple way of generating precisely controlled high-dimensional spectral qudits, with the potential of harnessing in wavelength-multiplexed quantum networks, high-dimensional information processing, and communication of quantum states specifically, and fiber-optic quantum remote sensing.

## 1. Introduction

The non-separable statistical relations between subsystems of a compound quantum system, also known as entanglement, are a curious aspect of quantum physics and have their roots in the development of quantum information and communication technologies. Quantum entanglement states with dimensionality greater than two, i.e., qudits, serve as a key resource for achieving higher secure key rates and circumventing scalability limitations compared to many-particle two-dimensional systems. The various photonic entangled degrees of freedom, such as polarization[1], orbital angular momentum[2,3], path[4], spatial[5,6], energy-time[7,8], and time-frequency[9-16] modes, can serve as conveyors of high-dimensional quantum information. Among these, access to higher-dimensional Hilbert space via the time-frequency degree of freedom offers distinct advantages, encoding multiple qubits per photon in time-frequency modes while providing enhanced immunity against errors and noise, and thus enhancing both the storage and robustness of quantum information systems. While frequency-bin qudit states can be used for many quantum information protocols, including computing, sensing, and networking, their compatibility with generation using integrated quantum photonic devices and real-world fiber-optic infrastructure makes it especially promising for emerging wavelength-division-multiplexed quantum networks[17] and distribution of quantum correlation over long distances, accelerating the transition of quantum technologies from lab to field.

Frequency-entangled biphoton quantum states can be generated by harnessing nonlinear optical processes, such as spontaneous parametric down-conversion (SPDC) or spontaneous four-wave mixing (SFWM), pumped by either CW or pulsed laser sources. Energy conservation between the interacting optical fields in a SPDC-based second-order nonlinear medium inherently leads to continuous spectral entanglement between the emitted biphotons. This continuous broadband spectral entanglement can be discretized into well-defined frequency-bin qudit states and demonstrated utilizing several physical platforms[15]. This goal can be achieved by incorporating SPDC together with an optical cavity[9,16,18-20], domain-engineered nonlinear crystals[14], cascaded electro-optic phase modulators with Fourier-transform pulse shapers[11], and quantum interference[21-24] techniques. Recently a nonlinear interferometer composed of multiple nonlinear crystals in series, along with linear spectral phase control, has been theoretically proposed to generate high-dimensional spectral qudits[25]. The generation of explicitly discrete-frequency quantum states that are not components of a broader continuous distribution was demonstrated for the first time in ref. 26. On-chip sources[10,15,27] based on SFWM in third-order nonlinear microresonators have enabled frequency-entangled biphoton emission with comb-like spectra[28], offering stable, low-cost, and scalable platforms suitable for mass-producible chip-scale quantum light. Nevertheless, many of the above physical platforms that generate frequency-bin qudit states are expected to incur optical losses from filtering, and microring resonator losses. These shortcomings limit the achievable biphoton generation rate, while the typically fixed architectures such as the fixed free spectral range of optical micro-resonators or the complex fabrication requirements of nonlinear crystals offer limited controllability and tunability, restricting the number of accessible parameters such as bin spacing, intra-bin entanglement, and dimensionality, and often involve complex experimental components. At a deeper level, it is desirable to gain fine-tuning control over the frequency degree of freedom by manipulating the biphoton joint spectrum both in amplitude and phase. Such control can be achieved by manipulating the time-domain structure of a biphoton wave packet via Fourier optical synthesis. This approach enables the generation of versatile ultrafast frequency-bin biphoton qudit states without sacrificing source brightness and without relying on complex fabrication processes such as domain engineering or cavity-based architectures, thereby expanding the accessible experimental parameters.

The two-dimensional Fourier-transform description of biphoton states has been experimentally verified[29-31], demonstrating that quantum correlations in the time and frequency domains are mutually inverse. The feat, which relies on time-frequency Fourier duality, enables the precise control and synthesis of arbitrary temporal or spectral modes of biphoton wave packets by shaping the amplitude and phase in the complementary domain. Building on this framework, a recent work[13] presented a proof-of-principle demonstration of a relatively simple and flexible way to synthesize biphoton time-bin modes via frequency-domain quantum optical synthesis using a bidirectional SPDC pumping scheme and varying crystal temperature. The same group subsequently extended this approach to control the spectral properties of biphotons by manipulating their temporal structure, introducing the concept of time-domain quantum optical synthesis[32]. In that work[32], the relative phase between biphoton joint temporal modes generated in each SPDC pumping varied by introducing a dispersive optical delay component, enabling shaping of the biphoton joint spectrum and the synthesis of two broadband joint spectral modes with fixed frequency spacing. In a departure from the above experiment on time-domain quantum optical synthesis, we add a nonlinear interferometer to manipulate each biphoton temporal mode independently, allowing the precise synthesis of discretely frequency-entangled qudits with bin spacings ranging from 12.5 GHz to 750 GHz, created without any spectrally selective filtering. We demonstrate full control over both the frequency-bin spacing and phase of the created states while maintaining a high biphoton generation rate, addressing the challenge of controlling spectral quantum correlations. By combining this approach with previously developed frequency-domain quantum optical synthesis, we achieve high control over both the temporal and spectral characteristics of biphoton states, which is essential for photon-based high-dimensional quantum information technologies. Furthermore, we transmit one of the entangled qudit pairs through 1.3 km of single-mode fiber (SMF) in a campus-scale two-node fiber network and verify the preservation of spectral correlations through coincidence detection, highlighting the immediate applicability of our approach to fiber-based wavelength-division-multiplexed quantum networks powered by entanglement.

## 2. Experimental setup and results

### *2.1 Generation and characterization of frequency qudits*

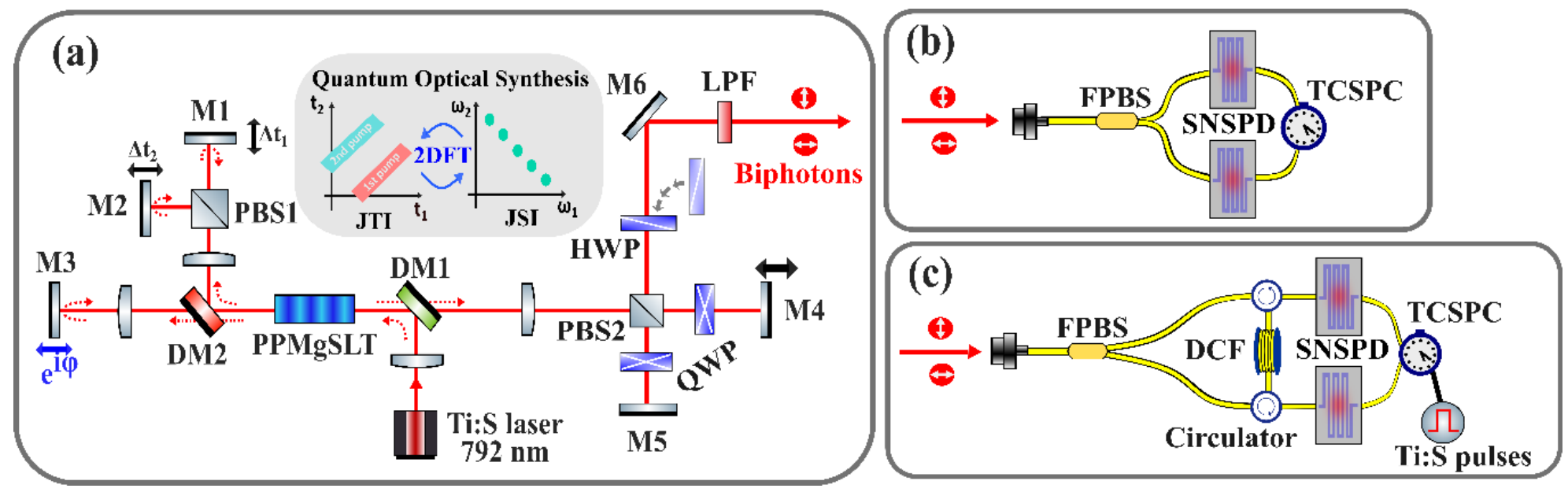

Fig.1 Schematic for generating and characterizing frequency-entangled biphotons: (a) Type-II SPDC-based PPMgSLT source of high-dimensional frequency-bin qudits from a Ti: Sapphire laser pulses through Fourier quantum optical synthesis. PPMgSLT: periodically poled magnesium oxide-doped stoichiometric lithium tantalate; DM: Dichroic mirror; M: Mirror; PBS: polarizing beamsplitter; QWP: Quarter-wave plate; HWP: Half-wave plate; LPF: Long pass filter. (b) Heralded coincidence measurement. FPBS: fiber polarizing beamsplitter; SNSPD: superconducting nanostrip single-photon detector; TCSPC: Time correlated single photon counting. (c) Joint spectral intensity (JSI) measurements with a fiber-based time-of-flight spectrometer. DCF: Dispersion compensating fiber module.

The experimental setup employed for the generation of frequency-bin qudits through time-domain Fourier quantum optical synthesis is illustrated in Fig. 1a. A mode-locked Ti:sapphire oscillator emits laser pulses at a repetition rate of 76 MHz, centered at 792 nm at a 3-dB bandwidth of 0.2 nm and steered to pump the SPDC biphoton source. The laser pulses, with an average power of 350 mW, are reflected at DM1 and then focused onto a periodically poled MgO-doped stoichiometric $LiTaO_3$ (PPMgSLT) nonlinear crystal having a poling period of 21.5 μm. The 40-mm-long PPMgSLT crystal is housed in a temperature-stabilized oven maintained at 42 °C to achieve a degenerate SPDC emission with a high biphoton generation rate. Because of its weak birefringence and thus small group-velocity mismatch between the constituent photons of biphotons, the type-II quasi-phase-matched PPMgSLT crystal facilitates broadband and frequency-anticorrelated biphotons[33,34] with orthogonal polarizations, $|H\rangle$ (denoted as signal) and $|V\rangle$ (idler).

We implement a bidirectional SPDC pumping scheme in which the pump light passes through the PPMgSLT crystal twice, with parametric down-conversion occurring during each pass, thereby providing two temporally separated wave packets[32]. Each pass results in a discrete joint temporal distribution exhibiting positive correlations, which are arranged and connected sequentially in time. For further temporal manipulation, a DM2 is placed between the PPMgSLT crystal and M3, directing the biphoton wave packet generated during the first pass into a nonlinear interferometer composed of a PBS1 and M1 and M2. Meanwhile, the pump light transmitted through DM2 is reflected back by M3 and subsequently passes through the PPMgSLT crystal, providing a second SPDC emission. By adjusting the positions of M1 and M2, independent temporal operations can be applied to each of the signal and idler photons emitted in the first pumping. It allows precise control over the temporal separation of the bidirectionally emitted two joint temporal distributions, including complete temporal overlap and/or controlled separation along the antidiagonal (Fig. 1a inset). As a result, two-photon interference in the frequency domain is induced, leading to the generation of arbitrary, well-separated frequency-bin entangled states. Consequently, a coherently superposed biphoton quantum state originating from the bidirectional pumping is obtained, given by $|\psi\rangle = (|2,0\rangle + e^{i\phi}|0,2\rangle)/\sqrt{2}$. Here, the first (second) term corresponds to the biphotons generated in the first (second) SPDC pass, while no biphotons were generated in the second (first) pass. The relative phase $\phi$ can be tuned by adjusting the position of M3, mounted on a piezo stage. Detailed mathematical derivations explaining the origin of frequency-bin state are provided in supplementary information. The biphotons generated in both passes co-propagate through DM1 and are directed into a modified polarization Michelson interferometer that replaces a conventional timing-compensation component. The Michelson interferometer, consisting of a PBS2, QWP, M4, and M5, compensates for the temporal walk-off between the orthogonally polarized biphotons. This configuration simultaneously helps us to tune the relative delay between the constituent photons, facilitating two-photon interference measurements. The residual pump field is filtered out with long-pass filters, and the time-compensated biphotons are coupled into a fiber-based polarization beam splitter. Polarization-separated biphotons are then detected using superconducting nano-strip single photon detectors (SNSPD), with the discriminated SNSPD electrical signals subsequently time-stamped by a time-interval analyzer for heralded coincidence measurements as shown in Fig. 1b. Our SNSPDs, with a timing jitter of 80 ps and a dark count rate of less than 100 counts per second, provide a detection efficiency of approximately 70%.

For a 341 mW of transmitted pump, we measure the singles $\sqrt{s_H s_V}$ of 1.970 million counts/second, and the coincidences $CC$ of 0.196 million counts/second in a window of 3 ns, operated below the saturation levels of our SNSPDs. Here $s_H$ and $s_V$ are the singles detected in the signal and idler arm, respectively, and $CC$ is the coincidences between them. Heralding detection yields a Klyshko efficiency $CC/\sqrt{s_H s_V}$ of (9.97 ± 0.16) % and measured pair generation rate of 587 ± 16 pairs/second per mW for transmitted pump powers ranging from 84 mW to 341 mW. Accounting for the Klyshko efficiency, which includes overall losses from scattering, optical transmission, fiber connectors, and SNSPD detection efficiency, the biphoton generation rate reaches ~59,000 pairs/second per mW. At an SPDC pump power of 1 W, the biphoton generation rate reaches on the order of tens of million photon pairs/second, highlighting its high generation efficiency.

Frequency-resolved joint spectral intensity (JSI) is directly measured using a simple and fast dispersive-fiber-based[13,29,31] time-of-flight method (Fig. 1c) that maps spectral correlations onto photon arrival times via a frequency-dependent group delay. The polarization separated $|H\rangle$ and $|V\rangle$ photons counter-propagate through the same dispersion-slope-compensating fiber module to ensure identical dispersion and thermal conditions, and coincidence arrival-time measurements relative to the SPDC pump pulses are converted to frequency scales to reconstruct the two-dimensional JSI distribution. The dispersion fiber module for nominal compensation of 50 km SMF has a typical insertion loss of 3.4 dB and a dispersion of -895 ps/nm at 1565 nm. Our spectrometer achieves a spectral resolution of approximately 0.14 nm with our total-system timing jitter of 130 ps, sufficient to fully resolve biphoton spectral correlations as shown in Fig. 2.

In order to modulate the spectral mode structure of biphotons via time-domain quantum-optical synthesis, the two temporally separated wave packets are manipulated in a controlled manner. First, we determine the temporal origin position, considered to be the condition in which all optical path lengths from DM2 to M1, M2, and M3 are equal. At this temporal origin, the two joint temporal distributions overlap, giving rise to parametric oscillations observed through two-photon interference measurements. Initially, the stages of the nonlinear interferometer are fixed at the position corresponding to the maximum change in coincidence counts in the interference measurement. Starting from this temporal origin, it becomes possible to tune overlapped joint temporal distributions aligned in parallel along the antidiagonal direction, as shown in Fig.1a inset, by adjusting the positions of M1 and M2 by equal amounts in opposite direction (eq. (S1.31)), i.e., moving one stage to shorten the optical path length relative to the temporal origin while moving the other stage to lengthen it. This temporal manipulation results in the formation of large range of frequency-bin states in the frequency domain. The inverse of the resulting temporal separation along the antidiagonal $\Delta t_-$ determines the corresponding frequency-bin spacing $\Delta\nu_-$ obeying time-frequency Fourier duality $\Delta\nu_- \times \Delta t_- = 1$ (eq. (S1.34)).

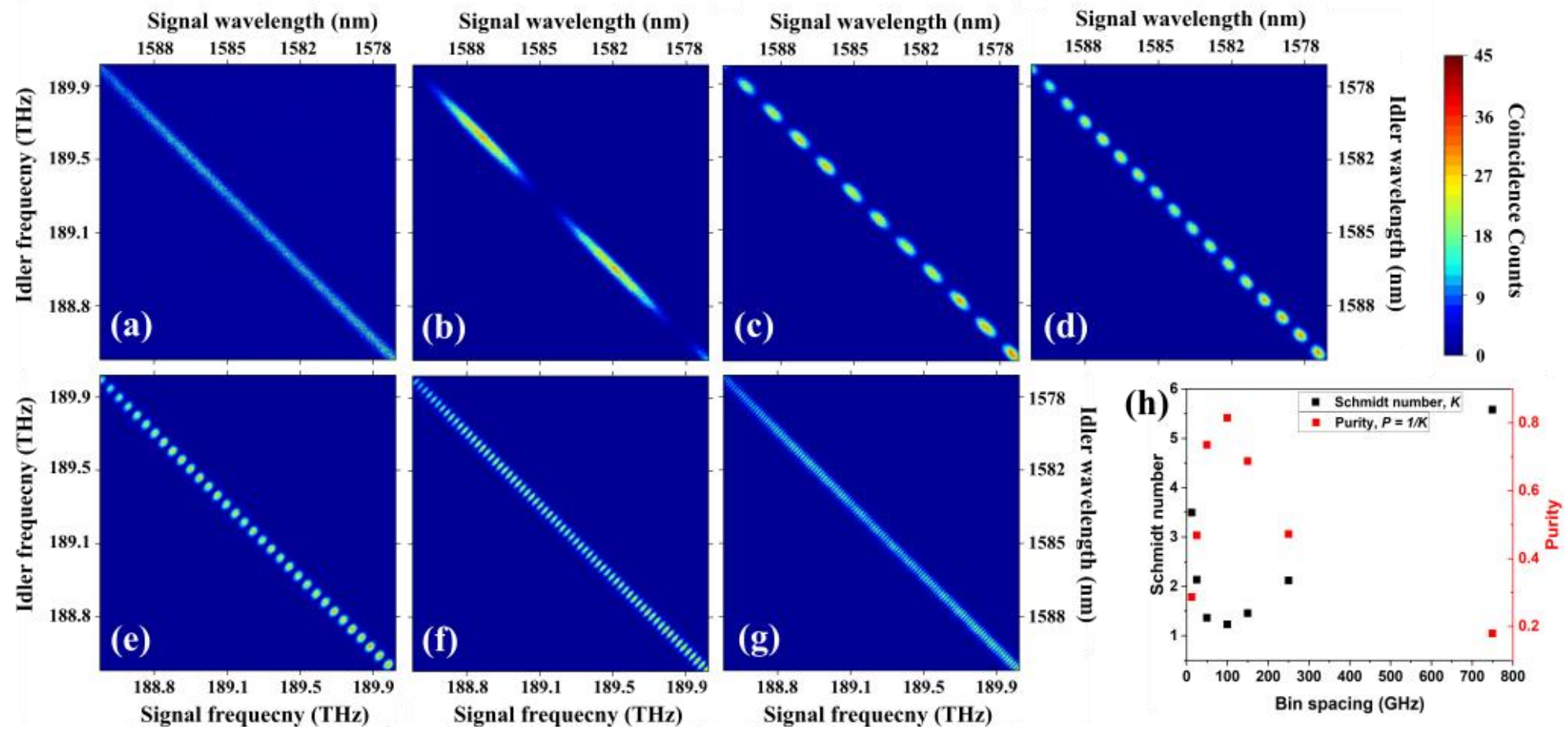

Fig.2 Frequency-resolved JSI distributions of biphotons at various frequency-bin spacings. (a) Continuous JSI of the frequency anticorrelated biphoton quantum state. The spectrum is then converted into discrete, well-defined frequency-bin structures with spacings of 750 GHz (b), 150 GHz (c), 100 GHz (d), 50 GHz (e), 25 GHz (f), and 12.5 GHz (g). Note that the X- and Y-axes span less than one pump pulse period due to the use of a longer DCF. (h) Corresponding biphoton dimensionality (Schmidt number) and purity of a single bin in each case.

First, we present the JSI measured with the interferometer stages fixed at the temporal origin. Figure 2a, acquired with an integration time of 120 s, showing photon coincidences only on the antidiagonal of the two-dimensional JSI matrix and revealing broadband, anticorrelated spectral correlations without any modulations. The relatively low count rate for this integration time is attributed to rather destructive nonlinear interference between the two temporal wave packets. Next, we present the frequency-bin states obtained after applying controlled temporal manipulation. Figures 2b-2g show the JSIs, each comprising a sequence of peaks with uniform width and equal spacing. Figure 2b is obtained when M1 and M2 in the nonlinear interferometer are displaced by +100 μm and −100 μm, respectively, while Fig. 2c corresponds to displacements of +500 μm and −500 μm. The resulting temporal spacing along the time-difference (antidiagonal) axis is determined through the nonunitary transformation $\Delta t_- = \Delta t_1 - \Delta t_2$. Converting these optical path lengths into relative temporal separations between the two biphoton wave packets yields temporal spacing of 1.33 ps for Fig. 2b and 6.67 ps for Fig. 2c on the marginal axes. We compare the expected relative temporal separation $\Delta t_-$ calculated from the applied optical path lengths, with the modulation period $\Delta\nu_-$ extracted from the projection of the measured JSI on the marginal axes. As observed, increasing the temporal separation between the two wave packets from 1.33 ps to 6.67 ps results in a shorter spectral modulation period, corresponding to frequency-bin spacing of 750 GHz (Fig. 2b) and 150 GHz (Fig. 2c), respectively, with the product of these two quantities approaching unity. From the above results, we confirm that the relative temporal separation $\Delta t_-$ between the two temporal wave packets and the spectral modulation period $\Delta\nu_-$ are in a Fourier-conjugate relationship. Next, making use of this property, we generate the frequency-bin states with bin spacings matched to the ITU grid (Figs. 2d-2g) used in wavelength-division multiplexing systems.

Temporal separations $\Delta t_{-}$ of 10 ps, 20 ps, 40 ps, and 80 ps yield frequency-bin spacings of 100 GHz (Fig. 2d), 50 GHz (Fig. 2e), 25 GHz (Fig. 2f), and 12.5 GHz (Fig. 2g), respectively, each acquired with an integration time of 60 s. For the 12.5 GHz case (Fig. 2g), the frequency-bin structure becomes more clearly resolved when a higher-dispersion DCF module is employed. To characterize the dimensionality of the generated frequency-bin states, we evaluate the Schmidt number, which represents the minimum number of relevant discrete orthonormal modes and thus quantifies the effective dimensionality of the biphoton state. We approximate the joint spectral amplitude by taking the square root of the measured JSI and perform Schmidt mode decomposition[10,31,35]. Figure 2h shows the Schmidt number $K$ evaluated for a single bin of the JSI as a function of frequency-bin spacing. As the bin spacing increases from 12.5 GHz to 100 GHz, the value of $K$ approaches unity, then increases again at larger spacings. This behavior stems from the evolution of the JSI distribution within each bin from an elliptical (correlated) to a circular (uncorrelated) distribution, and back to an elliptical (anticorrelated) form. In particular, the 100 GHz bin spacing yields intra-bin states with comparably high purity (~81%), where the JSI exhibits nearly circular, indicating separable and spectrally uncorrelated within each bin. Evaluating the full JSI including all the measured bins for the 100 GHz case (Fig. 2d) gives a dimensionality of $K \approx 17$, corresponding to a frequency-binned Hilbert space of at least $K \times K \approx 289$, formed by two entangled qudits with dimension $D = 17$. Note that the actual dimensionality is expected to be higher than 17, as the measured JSI spans less than one pump pulse period.

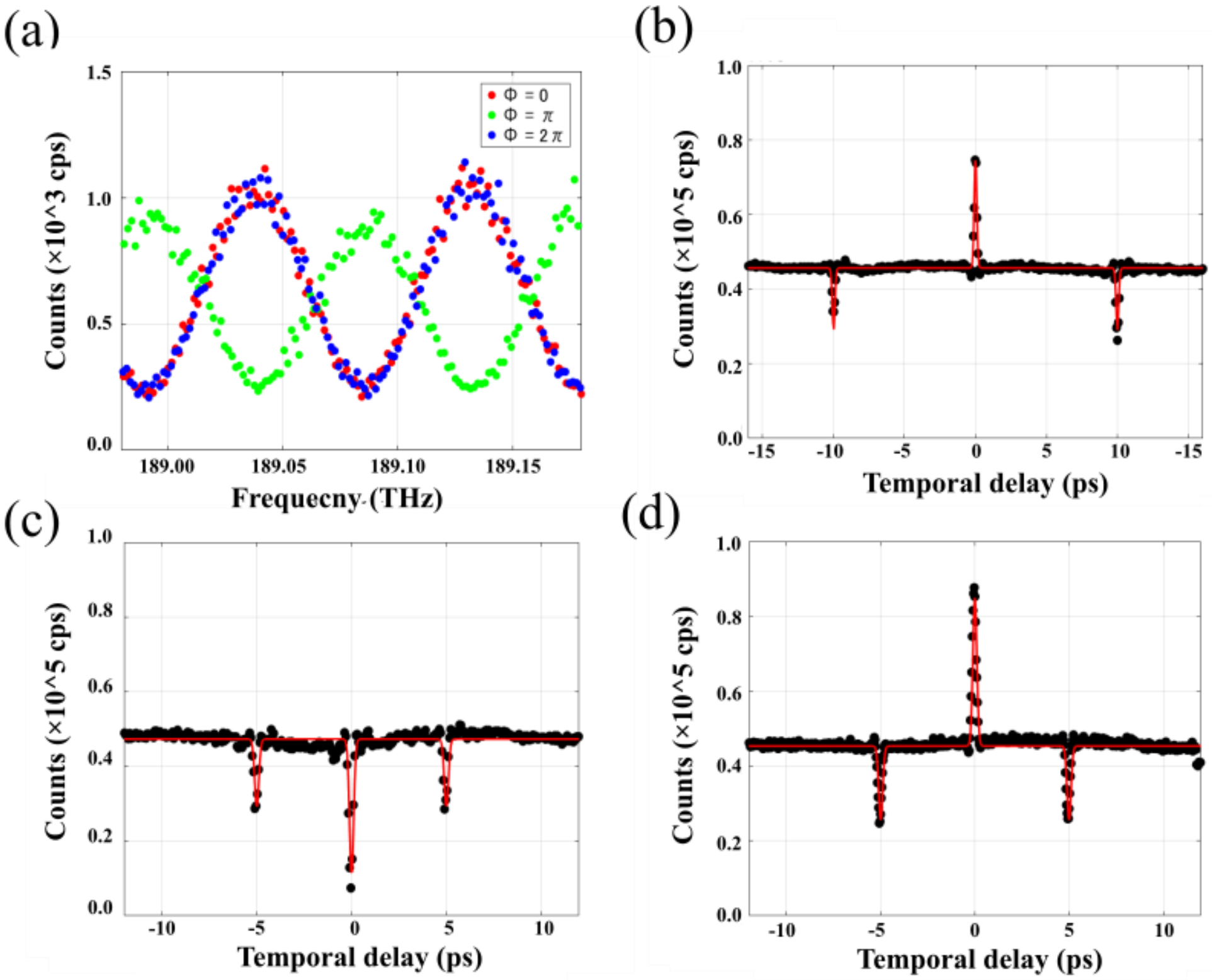

Fig. 3 Demonstration of spectral phase tuning of qudit states. (a) Marginal distributions of the JSI measured while tuning the relative phase from 0 (red) to π (green) and 2π (blue) by displacing M3 from the temporal origin to 200 nm and 400 nm. (b) HOM interference measured for a frequency-bin spacing of 50 GHz. (c) HOM interference measured at the temporal origin, and (d) when M3 is displaced by 200 nm, corresponding to a π phase shift for frequency-bin spacing of 100 GHz.

By changing the relative phase between the two temporal wave packets along the time-difference axis, it is possible to control the phase of the spectral modulation along the frequency-difference axis. In our experimental setup, the M3 mounted on a piezo stage allows tuning the relative phase $\phi$ between the two temporal modes precisely. An optical path length of 400 nm from the temporal origin, half of the wavelength of the pump light, induces a relative phase of $\pi$. As a result, it becomes possible to observe changes in the bright and dark positions in the biphoton joint spectrum. Figure 3a shows the marginal distributions of JSI measured at tuning the delay from 0 nm (red) to 400 nm (green) and 800 nm (blue). Flipping the spectral modulation between red (blue) and green data manifests the controllability of spectral phase modulation.

For further clarification, we present the estimation of the relative phase via Hong-Ou-Mandel (HOM) interference[36] measurement. We inserted HWP at 22.5° between PBS2 and M6 and measured coincidence counts as we scanned M4 on the motorized stage to observe HOM interference. The measured HOM interference results are shown in Figs. 3b-3d. Two side fringes and a central main fringe with a fringe visibility of 87% are observed. The two side dips arise from interference between identical temporal wave packets generated within the same pump pass, whereas the central fringe corresponds to interference between wave packets generated in the bidirectional pump passes. We investigate the relationship between the temporal interference fringe period and the frequency-bin spacing for 50 GHz and 100 GHz. The temporal fringe periods measured in Fig. 3b, and Figs. 3c and 3d are 9.98 ps and 4.89 ps, respectively. Accounting for the factor of two arising from the nonunitary transformation, these fringe periods correspond to frequency-bin spacings of 50 GHz and 100 GHz, in good agreement with the bin spacing of the generated frequency-bin qudit states. As the spectral modulation period increases, the temporal separation between the interference fringes correspondingly decreases. This correspondence indicates that discrete frequency entanglement manifests as oscillatory behavior in the temporal domain. Figure 3c shows the HOM interference pattern measured at the temporal origin, while Fig. 3d corresponds to the case in which M3 is displaced to introduce an optical path length of 400 nm. The clear inversion of the central fringe between Figs. 3c and 3d originates from a relative phase shift of $\pi$ between the two temporal wave packets. Together with the observed frequency anticorrelation and high-visibility temporal interference, these results confirm the generation of high-dimensional frequency-bin entanglement. By combining precise control of the frequency-bin spacing with tunable manipulation of the relative phase, we demonstrate and strengthen full quantum-regime spectral modulation, enabling a high degree of freedom through Fourier quantum-optical synthesis in two-dimensional time-frequency space, offering reconfigurability and significantly greater flexibility surpassing conventional methods for generating frequency bins.

### *2.2 Frequency entanglement distribution*

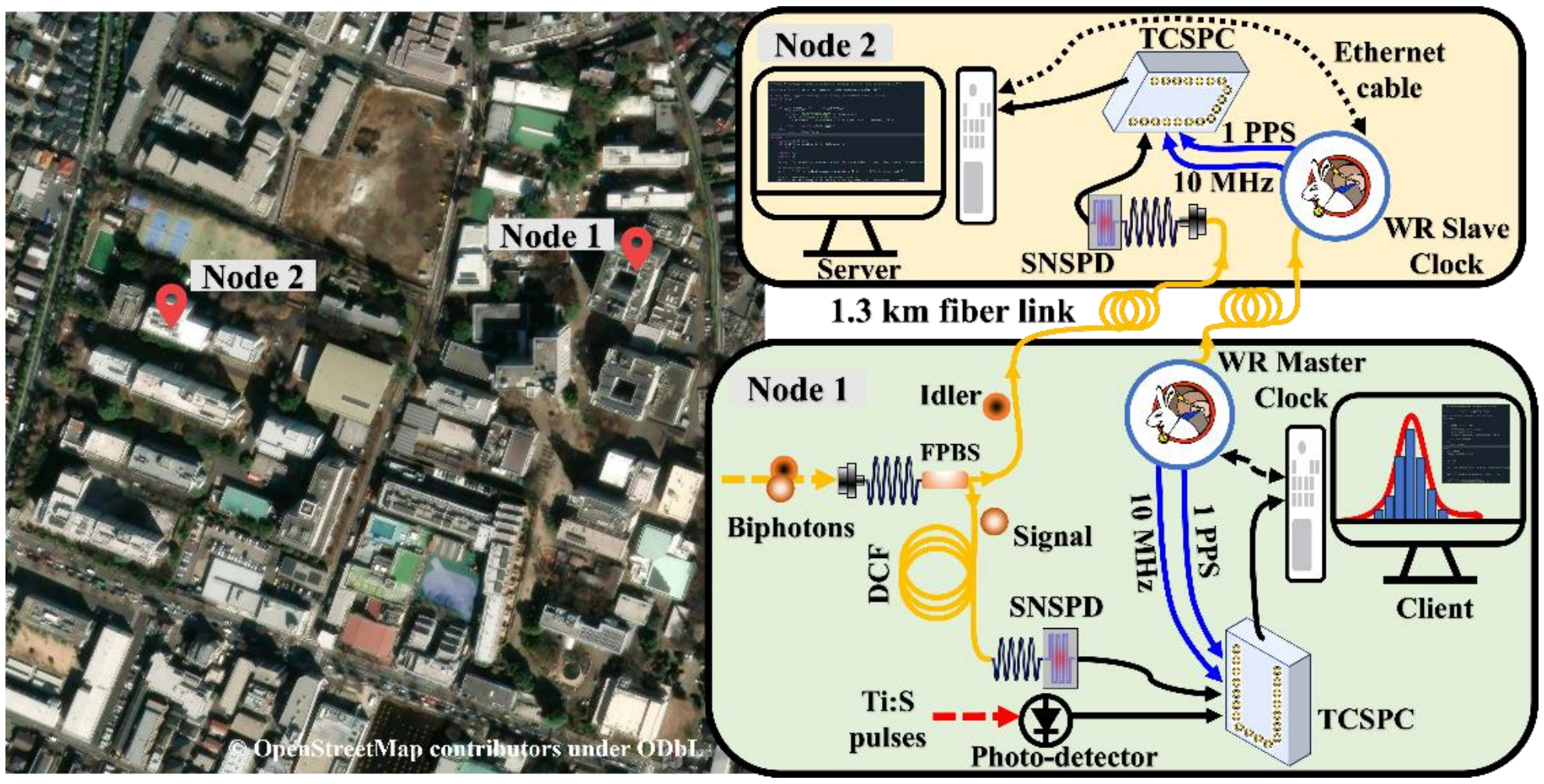

Fig. 4. Entanglement distribution network in our university campus. A visual representation of a campus-scale two-node optical fiber quantum network including highly precise time-synchronization between remote time taggers utilizing White Rabbit precision time protocol network, illustrating entanglement distribution over 1.3 km of fiber infrastructure. Map data ©Openstreetmap contributors. See[42] for WR logo copyright information.

A schematic of our two-node quantum network deployed across the east and west campuses of the University of Electro-Communications (UEC) is shown in Fig. 4. The network consists of two spatially separated nodes connected by approximately 1.3 km of single-mode optical fiber. The biphoton source is located at node 1 in east campus, where the generated photon pairs are split into signal and idler photons using an FPBS. The signal photons are transmitted through a DCF (nominal compensation of 15 km SMF) for spectral analysis and are detected locally using an SNSPD. Concurrently, the partner idler photons are transmitted to node 2 in west campus. In this experiment, the White Rabbit (WR) based classical timing signal from master WR LEN module (Safran) and the idler photons from the biphoton SPDC source are transmitted separately to node 2 via two parallel single-mode fiber channels deployed between the nodes. At node 2, the idler photons are directly detected by another SNSPD without spectral resolution. High timing precision is essential for verifying the distribution of entangled photon pairs. Therefore, the two quantum nodes are temporally synchronized via CERN-born open-source timing technology[37], WR precision time protocol network[38-40], ensuring two independently running time-tagging modules tick together with an accuracy of a fraction of a sub-nanosecond. The photon arrival times are recorded relative to the WR LEN time reference. The WR framework also supports simultaneous data transfer rates of 1.25 Gbps using the same fiber network via bidirectional small form-factor pluggable (SFP) transceiver modules. The detection events at each node are digitized using Time Tagger Ultra modules (Swabian Instruments), featuring an RMS timing jitter below 10 ps. The time-tagged data recorded at node 2 are transmitted in real time to node 1 via the WR-LEN classical Ethernet link, where a client interface executes time-correlation measurements. With the implemented WR clock distribution system achieving a timing precision of 28 ps over the 1.3 km link, the correlation histograms are subsequently sampled with a 30 ps bin width.

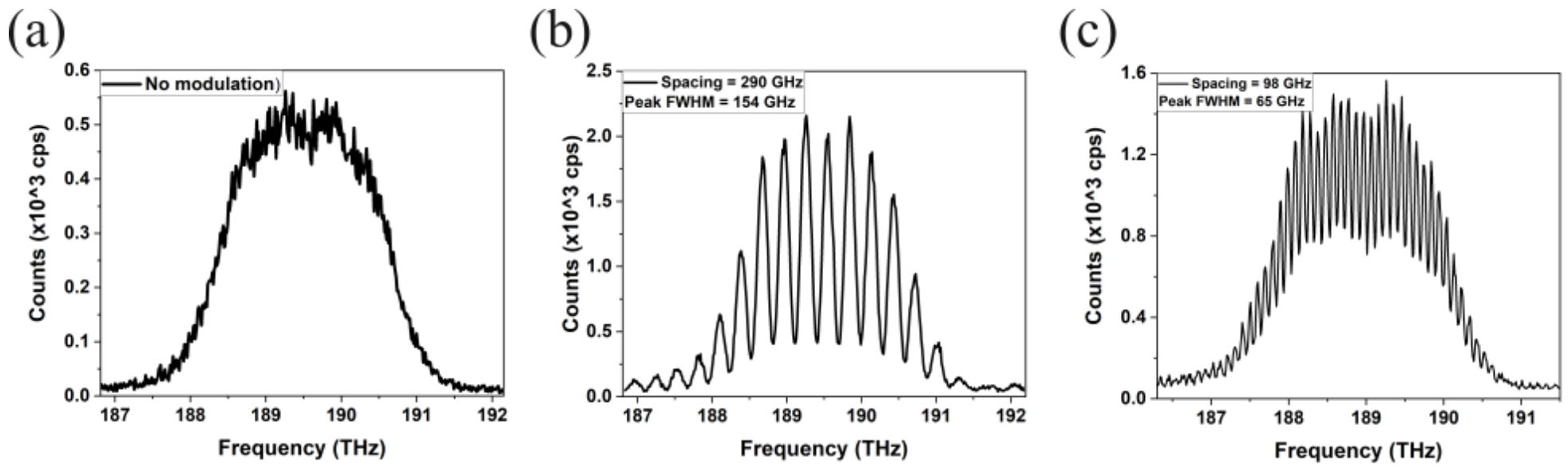

Fig. 5. Spectral distributions of the biphoton state after transmission through the fiber network. (a) Continuous spectral distribution without modulation. (b) Frequency-bin structure with a spacing of 290 GHz. (c) Frequency-bin structure with a spacing of 98 GHz. The frequency axis spans one pump pulse period, allowing up to 38 frequency bins to be clearly resolved in (c), counted at -10 dB level.

Frequency correlation between the remote and local photons is characterized via coincidence histograms as shown in Fig. 5, referenced to the pulse-train repetition rate. In this context, coincidences denote temporal correlations between local and remote photons, specifically those corresponding to time-stretched local photons. The histogram obtained in the absence of temporal modulation exhibits a continuous broadband spectral distribution of locally detected photons, as shown in Fig. 5a. Conversely, Fig. 5b shows the histogram with a frequency-bin spacing of 290 GHz, where regularly spaced Gaussian spectral peaks are observed across the envelope with a peak full width at half maximum (FWHM) of 154 GHz. By further increasing the temporal separation between the biphoton wave packets, a denser frequency-bin structure is obtained, as shown in Fig. 5c, with a bin spacing of 98 GHz and a peak FWHM of 65 GHz. The observation of 38 frequency-bin states within a single pump pulse period indicates that the actual dimensionality exceeds the previously estimated value of 17. The observation of these well-resolved periodic modulations with high-visibility quantum correlations confirms that discrete frequency-bin structures remain observable after transmission through the fiber link, demonstrating the feasibility of distributing high-dimensional frequency-encoded quantum states across a campus-scale network. Notably, when time-correlation measurements are executed without reference to the pulse-train repetition rate, the resulting spectra suffer significant degradation. This loss of resolution is attributed to the remote detector timing jitter and the chromatic dispersion accumulated over the 1.3 km fiber link. The WR-based distributed time-tagging architecture enables synchronized time-correlation measurements across distant nodes, supporting uninterrupted validation of non-local frequency correlations and entanglement distribution over a real-world fiber infrastructure. Furthermore, the network can be readily scaled by incorporating additional WR switch and WR LEN modules at each node, enabling high-precision synchronization across multiple nodes.

## 3. Summary

In summary, we have demonstrated a flexible method for engineering high-dimensional biphoton frequency-bin qudits via time-domain Fourier optical synthesis, addressing the challenge of precise, tunable control over spectral modulations. By using a nonlinear interferometer to independently manipulate the temporal structure of biphoton wave packets, we exploited time-frequency Fourier duality to achieve precise control over the spectral modes of quantum entangled optical states. We constructed discrete frequency-bin structures with spacings ranging from 12.5 GHz to 750 GHz, demonstrating direct compatibility with standard ITU grids used in classical telecommunications. Notably, at a frequency-bin spacing of 100 GHz, we achieved intra-bin pure states with a purity of 81% and measured frequency-binned Hilbert-space dimensionality of at least 289, corresponding to biphoton 17-dimensional qudit states. Total of 38 frequency-bins are observed within one pump pulse period, suggesting that the accessible Hilbert-space dimensionality is higher than the previously estimated value of 17. Our approach surpasses previous methods by enabling fully arbitrary modulation of the joint spectrum with a relatively simple and reconfigurable experimental setup. The spectral structure can be further modulated using the existing HOM interferometer with Fourier optical synthesis, resulting in the generation of finer sub-bin features within each frequency bin[23]. Furthermore, the distribution and verification of the generated frequency-bin qudits over a campus-scale fiber network highlights the potential of frequency-bin encoding for scalable wavelength-division-multiplexed quantum networks, where multiple frequency channels can be used to distribute high-dimensional quantum states across fiber-based infrastructure. Our results contribute to the global effort toward the WR-based synchronization of quantum networks enabling picosecond-level precision. Robust and ultra-high-precision synchronization in metropolitan-scale quantum networks is further feasible through quantum-classical coexistence transmission[40], which mitigates cumulative transmission time fluctuations. The ability to precisely engineer spectral qudits that meet the specifications of arrayed waveguide gratings opens exciting possibilities for establishing large-scale quantum networks.

Overall, this work establishes Fourier quantum optical synthesis as a versatile and viable method for active spectral engineering of high-dimensional photonic quantum states, achieving fine-grained control over bin spacing and phase of ultrafast frequency-bin qudits, a crucial element for time-frequency domain high-dimensional quantum information processing. Furthermore, we anticipate that this methodology will provide a framework for high-resolution remote quantum spectroscopy[41], and fiber-based quantum networks.

**Funding.** This work was supported by JSPS KAKENHI Grant Number JP 24K01379 and JSPS Program for Forming Japan's Peak Research Universities（J-PEAKS）Grant Number JPJS00420230003.

**Acknowledgments.** The authors acknowledge Kaoru Minoshima, Akifumi Asahara, and Masahiro Ishizeki of the University of Electro-Communications (UEC), Tokyo, Japan for their support in constructing the fiber-based quantum network on the UEC campus.

**Author contributions.** P.K.: Conceptualisation, Methodology, Data curation, Formal analysis, Investigation, Writing— Original Draft, Writing— review & editing. T.O.: Conceptualisation, Methodology, Data curation, Formal analysis, Investigation, Writing— review & editing. R.Y.: Supporting Data curation and analysis, and Methodology. Y.F.: Conceptualisation, Theoretical analysis, Supervision (supporting), Writing— Review & Editing. M.Y., H.T., S.M. & K.P.N.: Resources, Methodology (supporting). R.S.: Conceptualization, Funding acquisition, Methodology, Supervision (lead), Writing— review & editing.

**Disclosures.** The authors declare no conflicts of interest.

**Data availability.** Data underlying the results presented in this paper are not publicly available at this time but may be obtained from the authors upon reasonable request.

**Supplemental document.** See supplementary document for supporting theoretical framework.

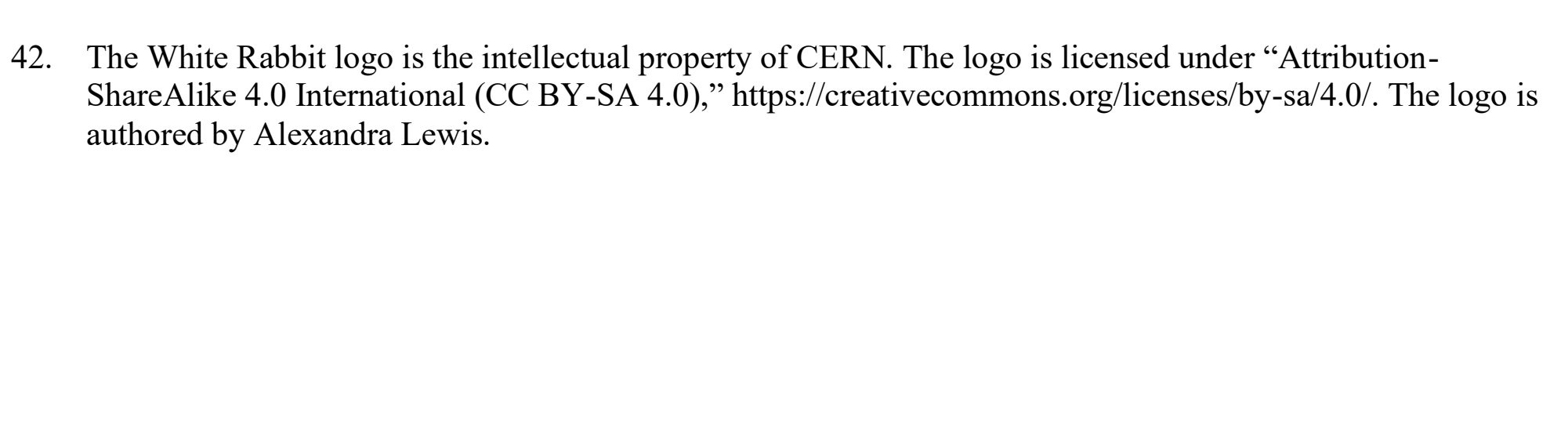

42. The White Rabbit logo is the intellectual property of CERN. The logo is licensed under “Attribution-ShareAlike 4.0 International (CC BY-SA 4.0),” https://creativecommons.org/licenses/by-sa/4.0/. The logo is authored by Alexandra Lewis.

# Supplementary document for "Quantum optical synthesis of high-dimensional ultrafast frequency-bin qudits"

**Prasad Koviri[1,†], Tomoya Okita[1,†], Rina Yabumoto[1], Yuta Fujihashi[1], Masahiro Yabuno[2], Hirotaka Terai[2], Shigehito Miki[2], Kali P. Nayak[1] & Ryosuke Shimizu[1,3*]**

*[1]Graduate School of Informatics and Engineering, The University of Electro-Communications, 1-5-1 Chofugaoka, Chofu, Tokyo 182-8585, Japan*
*[2]Advanced ICT Research Institute, National Institute of Information and Communications Technology, 588-2 Iwaoka, Nishi-ku, Kobe, Hyogo 651-2492, Japan*
*[3]Institute for Advanced Science, The University of Electro-Communications, 1-5-1 Chofugaoka, Chofu, Tokyo 182-8585, Japan*
*[†]These authors contributed equally to this work.*
**r-simizu@uec.ac.jp*

## 4. Theoretical description of bidirectionally pumped type-II SPDC

In this section, we present a derivation of the biphoton state generated in a bidirectionally pumped type-II spontaneous parametric down-conversion (SPDC) source. We consider a one-dimensional nonlinear crystal of length L. The two down-converted fields are orthogonally polarized and are labeled by horizontal (H) and vertical (V) polarization. The pump field is treated classically, whereas the generated fields are quantized.

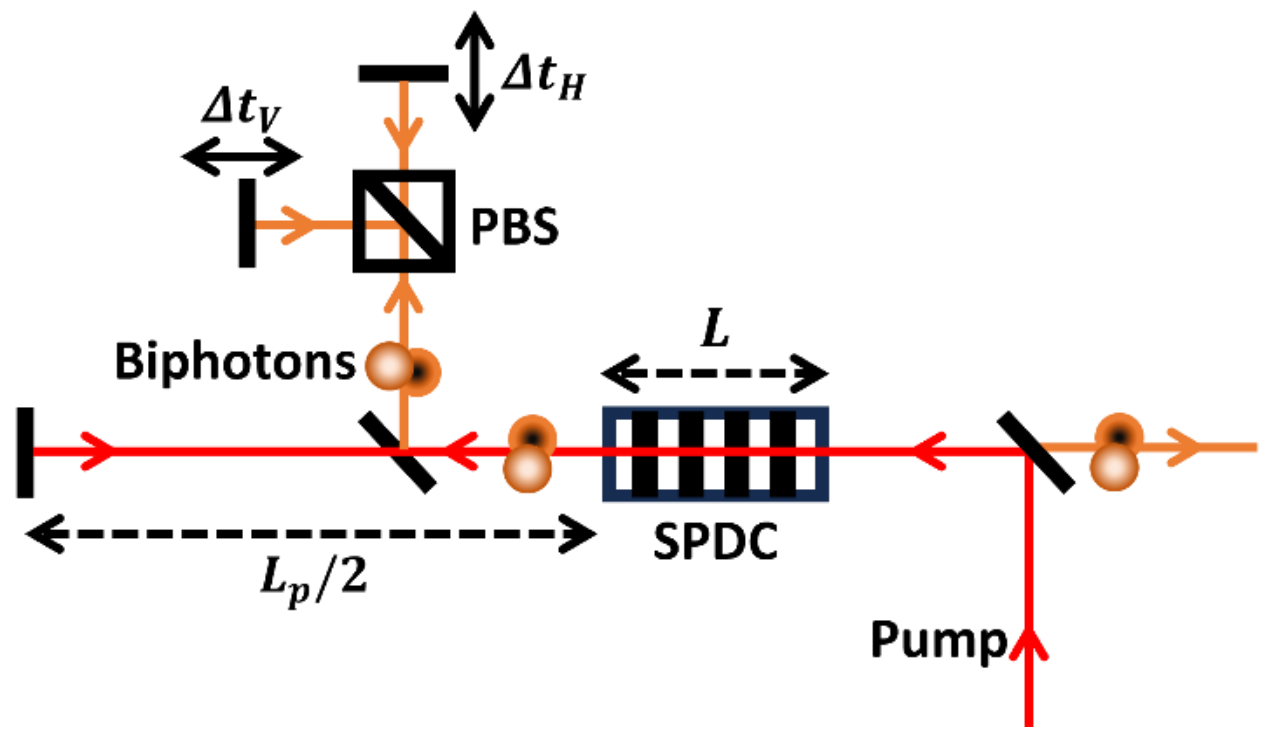

Fig.S1 Illustration for the time-domain quantum optical synthesis generating high-dimensional frequency-bin entangled biphotons. PBS: polarizing beamsplitter.

### A) Interaction Hamiltonian for the first and second passes

The interaction Hamiltonian[S1,S2] associated with the first pass of the pump through the crystal is written as

$$\hat{H}_{\mathrm{PDC1}}(t) = \int_{-\frac{L}{2}}^{\frac{L}{2}} dz\, \chi^{(2)} E_{\mathrm{p}}^{(+)}(z,t) \hat{E}_{1\mathrm{H}}^{(-)}(z,t) \hat{E}_{1\mathrm{V}}^{(-)}(z,t) + \mathrm{h.c.}, \tag{S1.1}$$

where

$$\hat{E}_{1\lambda}^{(-)}(z,t) = \left[\hat{E}_{1\lambda}^{(+)}(z,t)\right]^{\dagger} = \mathcal{C}_{\lambda} \int d\omega\, e^{-i[k_{\lambda}(\omega)z+\omega t]} \hat{a}_{1\lambda}^{\dagger}(\omega), \quad \lambda \in \{\mathrm{H},\mathrm{V}\}, \tag{S1.2}$$

and the classical pump field is expanded as

$$E_{\mathrm{p}}^{(+)}(z,t) = \mathcal{C}_{\mathrm{p}} \int d\omega_{\mathrm{p}}\, e^{i[k_{\mathrm{p}}(\omega_{\mathrm{p}})z+\omega_{\mathrm{p}} t]} \alpha(\omega_{\mathrm{p}}). \tag{S1.3}$$

Here $\alpha(\omega_{\mathrm{p}})$ denotes the pump spectral envelope. Substituting eqs. (S1.2) and (S1.3) into eq. (S1.1), we obtain

$$\begin{aligned} \hat{H}_{\mathrm{PDC1}}(t) &= \int_{-\frac{L}{2}}^{\frac{L}{2}} d\,z \int d\omega_a\, d\omega_b\, d\omega_{\mathrm{p}}\, \Gamma\, e^{-i[k_{\mathrm{H}}(\omega_a)+k_{\mathrm{V}}(\omega_b)-k_{\mathrm{p}}(\omega_{\mathrm{p}})]z} e^{-i(\omega_a+\omega_b-\omega_{\mathrm{p}})t} \\ &\quad \times \alpha(\omega_{\mathrm{p}}) \hat{a}_{1\mathrm{H}}^{\dagger}(\omega_a) \hat{a}_{1\mathrm{V}}^{\dagger}(\omega_b) + \mathrm{h.c.}, \end{aligned} \tag{S1.4}$$

where $\Gamma \equiv \chi^{(2)} \mathcal{C}_{\mathrm{p}} \mathcal{C}_{\mathrm{H}} \mathcal{C}_{\mathrm{V}}$ collects all frequency-independent prefactors.

The interaction Hamiltonian associated with the second pass has the same structure, but must additionally include the free-space propagation phases acquired by the pump and by the down-converted fields between the two passages through the crystal. It can therefore be written as

$$\hat{H}_{\mathrm{PDC2}}(t) = \int_{-\frac{L}{2}}^{\frac{L}{2}} d\,z \int d\omega_a\, d\omega_b\, d\omega_{\mathrm{p}}\, \Gamma\, e^{-i[k_{\mathrm{H}}(\omega_a)+k_{\mathrm{V}}(\omega_b)-k_{\mathrm{p}}(\omega_{\mathrm{p}})]z} e^{-i(\omega_a+\omega_b-\omega_{\mathrm{p}})t} \times e^{-\frac{i(\omega_a L_{\mathrm{H}}+\omega_b L_{\mathrm{V}}-\omega_{\mathrm{p}} L_{\mathrm{p}})}{c}} \alpha(\omega_{\mathrm{p}}) \hat{a}^\dagger_{2\mathrm{H}}(\omega_a) \hat{a}^\dagger_{2\mathrm{V}}(\omega_b) + \mathrm{h.c.}, \tag{S1.5}$$

where $L_{\mathrm{H}}$ and $L_{\mathrm{V}}$ are the free-space propagation lengths of the H- and V-polarized down-converted fields, respectively, and $L_{\mathrm{p}}$ is the free-space propagation length of the pump.

## B) Effective frequency-domain Hamiltonians

The unitary transformations generated by the first and second passes are

$$\hat{U}_{\mathrm{PDC1}} = \exp\left[-\frac{i}{\hbar}\int_{-\infty}^{\infty} dt\, \hat{H}_{\mathrm{PDC1}}(t)\right], \tag{S1.6}$$

$$\hat{U}_{\mathrm{PDC2}} = \exp\left[-\frac{i}{\hbar}\int_{-\infty}^{\infty} dt\, \hat{H}_{\mathrm{PDC2}}(t)\right]. \tag{S1.7}$$

We first evaluate the time integral for the first pass. Using

$$\int_{-\infty}^{\infty} dt\, e^{-i(\omega_a+\omega_b-\omega_{\mathrm{p}})t} = 2\pi\, \delta(\omega_a + \omega_b - \omega_{\mathrm{p}}), \tag{S1.8}$$

eq. (S1.4) gives

$$\int_{-\infty}^{\infty} dt\, \hat{H}_{\mathrm{PDC1}}(t) = 2\pi\Gamma \int_{-\frac{L}{2}}^{\frac{L}{2}} d\,z \int d\omega_a\, d\omega_b\, d\omega_{\mathrm{p}}\, e^{-i[k_{\mathrm{H}}(\omega_a)+k_{\mathrm{V}}(\omega_b)-k_{\mathrm{p}}(\omega_{\mathrm{p}})]z} \times \delta(\omega_a + \omega_b - \omega_{\mathrm{p}}) \alpha(\omega_{\mathrm{p}}) \hat{a}^\dagger_{1\mathrm{H}}(\omega_a) \hat{a}^\dagger_{1\mathrm{V}}(\omega_b) + \mathrm{h.c.} \tag{S1.9}$$

The integration over $\omega_{\mathrm{p}}$ can now be carried out immediately, which enforces energy conservation, $\omega_{\mathrm{p}} = \omega_a + \omega_b$. Defining the phase mismatch as

$$\Delta k(\omega_a, \omega_b) \equiv k_{\mathrm{H}}(\omega_a) + k_{\mathrm{V}}(\omega_b) - k_{\mathrm{p}}(\omega_a + \omega_b), \tag{S1.10}$$

we arrive at

$$\int_{-\infty}^{\infty} dt\, \hat{H}_{\mathrm{PDC1}}(t) = 2\pi\Gamma \int d\omega_a\, d\omega_b\, \alpha(\omega_a + \omega_b) \left[\int_{-\frac{L}{2}}^{\frac{L}{2}} dz\, e^{-i\Delta k(\omega_a,\omega_b)z}\right] \times \hat{a}^\dagger_{1\mathrm{H}}(\omega_a) \hat{a}^\dagger_{1\mathrm{V}}(\omega_b) + \mathrm{h.c.} \tag{S1.11}$$

The integration over $z$ gives the standard phase-matching function,

$$\int_{-\frac{L}{2}}^{\frac{L}{2}} dz\, e^{-i\Delta k z} = L\, \mathrm{sinc}\left(\frac{\Delta k L}{2}\right). \tag{S1.12}$$

Hence the first-pass unitary is expressed as

$$\hat{U}_{\mathrm{PDC1}} = \exp\left(-\frac{i}{\hbar}\hat{H}_{\mathrm{PDC1}}\right) \tag{S1.13}$$

with the effective Hamiltonian

$$\hat{H}_{\mathrm{PDC1}} = A\int d\omega_a\, d\omega_b\; f(\omega_a,\omega_b)\hat{a}_{1\mathrm{H}}^{\dagger}(\omega_a)\hat{a}_{1\mathrm{V}}^{\dagger}(\omega_b) + \mathrm{h.c.}, \tag{S1.14}$$

where all overall constants have been absorbed into a single coupling constant $A$, and

$$f(\omega_a,\omega_b) = \alpha(\omega_a+\omega_b)\mathrm{sinc}\left[\frac{\Delta k(\omega_a,\omega_b)L}{2}\right] \tag{S1.15}$$

is the joint spectral amplitude (JSA).

The derivation for the second pass is completely analogous. After using Eq. (S1.8), the extra free-space phase in eq. (S1.5) becomes

$$e^{-\frac{i(\omega_a L_{\mathrm{H}}+\omega_b L_{\mathrm{V}}-\omega_{\mathrm{p}}L_{\mathrm{p}})}{c}} \to e^{-\frac{i[\omega_a(L_{\mathrm{H}}-L_{\mathrm{p}})+\omega_b(L_{\mathrm{V}}-L_{\mathrm{p}})]}{c}}, \tag{S1.16}$$

so that

$$\hat{U}_{\mathrm{PDC2}} = \exp\left(-\frac{i}{\hbar}\hat{H}_{\mathrm{PDC2}}\right) \tag{S1.17}$$

with

$$\hat{H}_{\mathrm{PDC2}} = A\int d\omega_a\, d\omega_b\; e^{-\frac{i[\omega_a(L_{\mathrm{H}}-L_{\mathrm{p}})+\omega_b(L_{\mathrm{V}}-L_{\mathrm{p}})]}{c}} f(\omega_a,\omega_b)\hat{a}_{2\mathrm{H}}^{\dagger}(\omega_a)\hat{a}_{2\mathrm{V}}^{\dagger}(\omega_b) + \mathrm{h.c.} \tag{S1.18}$$

### C) Biphoton state in the low-gain regime

The state generated by the bidirectionally pumped source is

$$|\psi_{\mathrm{PDC}}\rangle = \hat{U}_{\mathrm{PDC2}}\hat{U}_{\mathrm{PDC1}}|\mathrm{vac}\rangle. \tag{S1.19}$$

In the low-gain regime we retain only terms linear in the SPDC efficiency. Expanding the unitary operators to first order gives

$$\hat{U}_{\mathrm{PDC1}} \simeq 1 - \frac{i}{\hbar}\hat{H}_{\mathrm{PDC1}}, \tag{S1.20}$$

$$\hat{U}_{\mathrm{PDC2}} \simeq 1 - \frac{i}{\hbar}\hat{H}_{\mathrm{PDC2}}. \tag{S1.21}$$

Substituting eqs. (S1.20) and (S1.21) into eq. (S1.19), and dropping terms of order $A^2$ and higher, we find

$$\begin{aligned} |\psi_{\mathrm{PDC}}\rangle &\simeq \left(1-\frac{i}{\hbar}\hat{H}_{\mathrm{PDC2}}\right)\left(1-\frac{i}{\hbar}\hat{H}_{\mathrm{PDC1}}\right)|\mathrm{vac}\rangle \\ &\simeq |\mathrm{vac}\rangle - \frac{i}{\hbar}\hat{H}_{\mathrm{PDC1}}|\mathrm{vac}\rangle - \frac{i}{\hbar}\hat{H}_{\mathrm{PDC2}}|\mathrm{vac}\rangle. \end{aligned} \tag{S1.22}$$

Because the Hermitian-conjugate terms annihilate the vacuum, only the creation operators contribute, so that

$$
\begin{aligned}
|\psi_{\mathrm{PDC}}\rangle \;\; &\simeq |\mathrm{vac}\rangle - \frac{iA}{\hbar}\int d\omega_a\, d\omega_b\; f(\omega_a,\omega_b)\hat{a}^{\dagger}_{1\mathrm{H}}(\omega_a)\hat{a}^{\dagger}_{1\mathrm{V}}(\omega_b)|\mathrm{vac}\rangle \\
&- \frac{iA}{\hbar}\int d\omega_a\, d\omega_b\; e^{-\frac{i[\omega_a(L_{\mathrm{H}}-L_{\mathrm{p}})+\omega_b(L_{\mathrm{V}}-L_{\mathrm{p}})]}{c}} f(\omega_a,\omega_b)\hat{a}^{\dagger}_{2\mathrm{H}}(\omega_a)\hat{a}^{\dagger}_{2\mathrm{V}}(\omega_b)|\mathrm{vac}\rangle .
\end{aligned}
\tag{S1.23}
$$

When the optical paths are aligned such that the spatial modes generated in the first and second passes are perfectly matched, the corresponding creation operators may be identified,

$$
\hat{a}^{\dagger}_{2\mathrm{H}}(\omega) = \hat{a}^{\dagger}_{1\mathrm{H}}(\omega), \qquad \hat{a}^{\dagger}_{2\mathrm{V}}(\omega) = \hat{a}^{\dagger}_{1\mathrm{V}}(\omega). \tag{S1.24}
$$

Equation (S1.23) then reduces to

$$
|\psi_{\mathrm{PDC}}\rangle = |\mathrm{vac}\rangle - \frac{iA}{\hbar}\int d\omega_a\, d\omega_b\, \left[1 + e^{-i(\omega_a \Delta t_{\mathrm{H}} + \omega_b \Delta t_{\mathrm{V}})}\right] f(\omega_a,\omega_b)\hat{a}^{\dagger}_{1\mathrm{H}}(\omega_a)\hat{a}^{\dagger}_{1\mathrm{V}}(\omega_b)|\mathrm{vac}\rangle , \tag{S1.25}
$$

where

$$
\Delta t_{\mathrm{H}} = \frac{L_{\mathrm{H}} - L_{\mathrm{p}}}{c}, \qquad \Delta t_{\mathrm{V}} = \frac{L_{\mathrm{V}} - L_{\mathrm{p}}}{c}. \tag{S1.26}
$$

Equation (S1.25) shows that bidirectional pumping modifies the JSA by the interference factor $1 + e^{-i(\omega_a \Delta t_{\mathrm{H}} + \omega_b \Delta t_{\mathrm{V}})}$.

### D) Time-domain representation

To obtain the corresponding joint temporal amplitude (JTA), we use the Fourier transform

$$
\hat{a}^{\dagger}_{1\lambda}(\omega) = \frac{1}{\sqrt{2\pi}}\int dt\, e^{i\omega t}\hat{a}^{\dagger}_{1\lambda}(t)\,, \quad \lambda \in \{\mathrm{H},\mathrm{V}\}, \tag{S1.27}
$$

and define

$$
f(t_a,t_b) = \frac{1}{2\pi}\int d\omega_a\, d\omega_b\, e^{i\omega_a t_a} e^{i\omega_b t_b} f(\omega_a,\omega_b)\,. \tag{S1.28}
$$

Substituting eq. (S1.27) into eq. (S1.25), the term proportional to the extra phase becomes

$$
e^{i\omega_a t_a} e^{i\omega_b t_b} e^{-i(\omega_a \Delta t_{\mathrm{H}} + \omega_b \Delta t_{\mathrm{V}})} = e^{i\omega_a (t_a - \Delta t_{\mathrm{H}})} e^{i\omega_b (t_b - \Delta t_{\mathrm{V}})}, \tag{S1.29}
$$

which directly implies a temporal shift of the JTA. The biphoton state can therefore be written as

$$
|\psi_{\mathrm{PDC}}\rangle = |\mathrm{vac}\rangle - \frac{iA}{\hbar}\int dt_a\, dt_b\, [f(t_a,t_b) + f(t_a - \Delta t_{\mathrm{H}}, t_b - \Delta t_{\mathrm{V}})]\hat{a}^{\dagger}_{1\mathrm{H}}(t_a)\hat{a}^{\dagger}_{1\mathrm{V}}(t_b)|\mathrm{vac}\rangle . \tag{S1.30}
$$

Thus, in the time domain, bidirectional pumping produces a coherent superposition of the original JTA and a delayed copy of it.

### E) Frequency-bin structures in the joint spectral intensity

Equation (S1.25) clarifies how a discrete frequency-bin structure emerges in the joint spectrum. In particular, when the temporal shifts are applied in opposite directions, i.e.,

$$\Delta t_{\mathrm{H}} = -\Delta t_{\mathrm{V}}, \tag{S1.31}$$

the phase factor in eq. (S1.25) depends only on the frequency difference $\omega_a - \omega_b$. The observable joint spectral intensity is then proportional to

$$\begin{aligned} |f(\omega_a, \omega_b)|^2 \left|1 + e^{-i(\omega_a \Delta t_{\mathrm{H}} + \omega_b \Delta t_{\mathrm{V}})}\right|^2 &= |f(\omega_a, \omega_b)|^2 \left|1 + e^{-i(\omega_a - \omega_b)\Delta t_{\mathrm{H}}}\right|^2 \\ &= 2|f(\omega_a, \omega_b)|^2 \big[1 + \cos\big((\omega_a - \omega_b)\Delta t_{\mathrm{H}}\big)\big]. \end{aligned} \tag{S1.32}$$

Hence, the modulation appears only along the frequency-difference axis, while the smooth spectral envelope is still determined by the function $f(\omega_a, \omega_b)$ defined in eq. (S1.25). The maxima satisfy

$$(\omega_a - \omega_b)\Delta t_{\mathrm{H}} = 2\pi m, \qquad m \in \mathbb{Z}, \tag{S1.33}$$

so that the spacing between adjacent maxima is given by

$$\Delta(\omega_a - \omega_b) = \frac{2\pi}{|\Delta t_{\mathrm{H}}|}, \qquad \Delta\nu = \frac{1}{|\Delta t_{\mathrm{H}}|}. \tag{S1.34}$$

Therefore, increasing the temporal separation in the antidiagonal direction decreases the spacing of the spectral modulation, and the continuous broadband joint spectrum is converted into an evenly spaced frequency-bin structure. This is the origin of the frequency-bin patterns observed experimentally in Fig. 2 of the main text.